\documentclass[aps,pra,nofootinbib,preprint,amsmath,amssymb,floatfix]{revtex4-2}
\usepackage{color}
\usepackage{graphicx}
\usepackage{mathtools}

\begin{document}

\newcommand{\tc}{\textcolor}
\newcommand{\g}{blue}
\title{On the Axion  Electrodynamics in a two-dimensional slab and the Casimir effect }
        % Enter your title between curly braces
{\author{ Iver H. Brevik  }      % Enter your name between curly braces
\affiliation{Department of Energy and Process Engineering, Norwegian University of Science and Technology, N-7491 Trondheim, Norway}
\author{Moshe M. Chaichian}
\affiliation{Department of Physics, University of Helsinki, and Helsinki Institute of Physics,  P. O. Box 64, FI-00014 Helsinki, Finland}
\author{Amedeo M. Favitta}
\affiliation{ Dipartimento di Fisica e Chimica—Emilio Segrè, Università degli Studi di Palermo, Via Archirafi 36, I-90123 Palermo, Italy }

\date{\today \\ aaTo be published in the International Journal of Modern Physics A}          % Enter your date or \today between curly braces

%\maketitle
%\affiliation

\begin{abstract}

We analyze the Axion Electrodynamics in a two-dimensional slab of finite width $L$ containing  a homogeneous and isotropic dielectric medium with constant permittivity and permeability.  We  start from the known decomposition of modes in the nonaxion case and then solve perturbatively the governing equations for the electromagnetic fields to which the axions are also coupled. This is a natural approach, since  the finiteness of $L$ destroys the spatial invariance of the theory in the $z$ direction normal to the plates. In this way we derive the value of the axion-generated rotation angle of the electric and magnetic fields after their passage through the slab, and use the obtained results to calculate the Casimir force between the two conducting plates. Our calculations make use of the same method as previously outlined in \cite{hoye20} for the case of Casimir calculations in chiral media and extend former results on the Casimir force in Axion Electrodynamics.
\end{abstract}
\maketitle

\bigskip
\section{Introduction}
\label{secintro}

The concept of axion, as a pseudoscalar particle, can be introduced in at least two different ways, either as a remedy to partially explain the CP problem in high energy physics - this was the original approach pioneered by Peccei and Quinn back in 1977 \cite{peccei77, peccei08} - or it can be introduced as a formal extension of the electrodynamics in the sense that one of the field tensors in the Lagrangian is replaced by its dual one. We will here follow the latter approach, although the two are compatible. Our intention is to discuss some fundamental issues of the formalism assuming a very simple geometric setup, namely a dielectric slab of finite width $L$ in the $z$ direction, with the plates in the transverse $x$ and $y$ directions being infinite and  metallic. We assume that the permittivity $\varepsilon$ and the permeability $\mu$ are constants, the constitutive relations in the rest system being ${\bf D}=\varepsilon {\bf E}, \, {\bf B}=\mu {\bf H}$. To summarize:

\bigskip

\noindent 1. Starting from the known electromagnetic stationary TE and TM modes in the cavity in the nonaxionic case, we discuss how the axions modify the formalism in the limit $L\rightarrow \infty$. This case is easy to describe, since the translational invariance in the $z$ direction allows us to introduce a wave vector component $k_z$ in this direction. In turn, this  makes it possible to derive the dispersive relations in a simple way. We consider two different cases for the pseudoscalar axion field $a$, either that its derivative with respect to $t$  is constant, $\dot{a}= \alpha$, or that its space derivative is constant, $da/dz=\beta$. In principle, both these requirements can be satisfied  at the same time, but in practice only one of them usually occurs. The space derivative of $a$ is generally associated with a three-dimensional force, while the time derivative of $a$ is associated with an interchange of energy.

\noindent 2. We next consider the governing equations for the electric and magnetic fields when the width $L$ is finite. The absence of translational invariance in the $z$ direction causes us to express the variations in this direction as derivatives with respect to $z$ directly. We take the influence from the axions to be small, and solve the inhomogeneous governing equations perturbatively to the first order in the nondimensional coupling parameter (to be specified later).

\noindent 3. The solutions show that the transverse fields $\bf E$ and $\bf H$ rotate slowly during the propagation in the slab. This makes it possible to calculate the Casimir force between the metal plates, using the rotation angle as input in the standard expression for the Casimir force for a nonaxionic medium between two metal plates. The calculation is given for finite temperatures. This method of calculating the Casimir force was worked out in an earlier paper for chiral media \cite{hoye20} (see also an earlier related work in \cite{jiang19}).

By now research on different aspects of axion physics is not new.
 Some of the pioneering papers on axions and the Axion Electrodynamics are listed in Refs.~\cite{peccei77,peccei08,sikivie83,weinberg78,preskill83,abbott83,dine83,carroll90,sikivie08,sikivie14}. More recent works can be found in Refs.~\cite{eggemeier23,millar17,liu22,li91,lawson19,kim19,qingdong19,sikivie03,mcdonald20,chaichian20,zyla20,arza20,carenza20,leroy20,brevik20,brevik21a,oullet19,arza19,
qiu17,dror21,brevik22a,fukushima19,tobar19,bae22,adshead20,patkos22,tobar22,derocco18,brevik22,brevik23,favitta23}.

In this paper we use the units as $\hbar = c = 1$.

\section{Fundamental modes in the nonaxionic case}

For reference purposes, it  is desirable first to summarize the known electromagnetic stationary modes in the cavity described above. In the nonaxionic case we can represent the wave vector $\bf k$ in terms of its three components, ${\bf k}=(k_x,k_y,k_z)$. For simplicity we will henceforth write $k$ instead of $k_z$. With $N$ a normalization constant we can express the TE mode as
\begin{align}
E_x=N\omega k_y\sin kz \,e^{i\Phi},  \nonumber \\
E_y = -N\omega k_x\sin kz \,e^{i\Phi}, \nonumber \\
B_x= -Nikk_x\cos kz \,e^{i\Phi}, \nonumber \\
B_y= -Nikk_y\cos kz \,e^{i\Phi}, \nonumber \\
B_z= -Nk_\perp^2\sin kz \,e^{i\Phi}, \label{TEmode}
\end{align}
where
\begin{equation}
\Phi= {\bf k}_\perp \cdot{\bf x}_\perp -\omega t, \quad k_\perp^2= k_x^2+k_y^2.
\end{equation}
The values of $k$ are discrete since the walls $z=0,L$ are conducting,
\begin{equation}
k= \frac{\pi p}{L}, \quad p= \pm 1, \pm 2,..., \label{p}
\end{equation}
and the dispersion relation in the cavity is
\begin{equation}
n^2\omega^2= k_\perp^2+k^2 = {\bf k}^2,
\end{equation}
with $n=\sqrt{\varepsilon \mu}$ the refractive index.

For the TM mode we have analogously
\begin{align}
E_x=N'ikk_x\sin kz \,e^{i\Phi}, \nonumber \\
E_y= N'ikk_y\sin kz \,e^{i\Phi}, \nonumber \\
E_z= -N'k_\perp^2\cos kz \,e^{i\Phi}, \nonumber \\
B_x=-N'\omega k_y\cos kz \,e^{i\Phi}, \nonumber \\
B_y= N'\omega k_x\cos kz \,e^{i\Phi}.
\end{align}
%It is of interest to relate the normalization constant $N$ to the energy density $W$ in the cavity. In general, in complex representation,
%\begin{equation}
%W= \frac{\varepsilon}{4}|{\bf E}|^2 + \frac{1}{4\mu}|{\bf B}|^2.
%\end{equation}
%For the TE mode we then get
%\begin{equation}
%W_{\rm TE} = \frac{k_\perp^2}{4\mu}
%N^2 (k^2+
%+2k_\perp^2\sin^2 kz),
%\end{equation}
%and for the TM mode,
%\begin{equation}
%W_{\rm TM}= \frac{\varepsilon k_\perp^2}{4}N^2( k^2+ 2k_\perp^2\cos^2 kz),
%\end{equation}
%with $k= \pi p/L$. Take the sum to get the total energy density in the cavity,
%\begin{equation}
%W= W_{\rm TE}+W_{TM} = \frac{1}{2}N^2k_\perp^2\varepsilon \omega^2;
%\end{equation}
%the $z$ dependent terms have gone away. This is the relation that connects the normalization constant $N$ with $W$.
\section{Basic equations}

For definiteness, we follow the conventions of Ref.~\cite{brevik23} when writing down the field equations. Thus we choose the metric convention $g_{00}= -1$ and introduce two field tensors $F_{\alpha\beta}$ and $H^{\alpha\beta}$, with $\alpha,\beta$ running from 0 to 3. The components of the original field tensor $F_{\alpha\beta}$ are as in vacuum, $F_{0i}=-E_i, F_{ik}= B_l$  where $i,k,l$ are cyclic, while the components of the contravariant response tensor $H^{\alpha\beta}$ are  $ H^{0i}=D_i, H^{ik}=H_l$.

We consider a pseudoscalar axion field $a= a({\bf r}, t)$  in the  universe,  making a two-photon interaction with the electromagnetic field.

The Lagrangian is \begin{equation}
{\cal{L}}= -\frac{1}{4}F_{\alpha\beta}{H}^{\alpha\beta} +{\bf A \cdot J}-\rho \Phi -\frac{1}{2} \partial_\mu a\partial^\mu a-\frac{1}{2}m_a^2a^2       - \frac{1}{4}g_{\alpha\gamma\gamma}a(x) F_{\alpha\beta}\tilde{F}^{\alpha\beta}. \label{1}
\end{equation}
in which the axion-two-photon coupling constant can be re-expressed as
\begin{equation}
g_{a\gamma\gamma}= g_\gamma \frac{\alpha}{\pi}\frac{1}{f_a}.
\end{equation}
Here $g_\gamma$, a  model-dependent constant, is usually taken to be $ 0.36$, as in the Dine-Fischler-Srednicki-Zhitnitski (DFSZ) invisible axion model \cite{Zhitnitsky:1980tq,DINE1981199,sikivie03}, or $-0.97$ as in the  Kim-Shifman-Vainshtein-Zakharov (KSVZ) model \cite{PhysRevLett.43.103,SHIFMAN1980493,sikivie03} ;  $\alpha$ is the fine structure constant, and $f_a$ is the axion decay constant. It is often assumed that $ f_a   \sim 10^{9}-10^{12}~$GeV \cite{sikivie08}.
The last term in the Lagrangian (\ref{1}) can be written  as ${\cal{L}}_{a\gamma\gamma} =  g_{a\gamma\gamma} a(x)\,{\bf E\cdot B}.$

We will henceforth make use of the nondimensional quantity $\theta(x)$, defined as
\begin{equation}
\theta(x)= g_{a\gamma\gamma}a(x).
\end{equation}
From the  expression (\ref{1}), one can derive the generalized Maxwell equations
\begin{equation}
{\bf \nabla \cdot D}= \rho-{\bf B\cdot \nabla}\theta, \label{5}
\end{equation}
\begin{equation}
{\bf \nabla \times H}= {\bf J}+\dot{\bf D}+\dot{\theta}{\bf B}+{\bf \nabla}\theta\times {\bf E}, \label{7}
\end{equation}
\begin{equation}
{\bf \nabla \cdot B}=0, \label{8}
\end{equation}
\begin{equation}
{\bf \nabla \times E} = -\dot{\bf B}. \label{9}
\end{equation}
It is to be noticed that these equations are general in the sense that  there are    no restrictions   on the spacetime variation of $a(x)$. The equations are moreover relativistic covariant with respect to shift of the inertial system. The latter property is  nontrivial, since the constitutive relations given above keep their simple form ${\bf D}=\varepsilon {\bf E}, \, {\bf B}=\mu {\bf H}$ in the rest system only. Covariance of the electromagnetic formalism is achieved by the introduction of {\it two} different field tensors, $F_{\mu\nu}$ and $H^{\mu\nu}$.

The  governing equations for the fields are
\begin{equation}
\nabla^2 {\bf E}-\varepsilon\mu \ddot{\bf E}=    {\bf \nabla (\nabla \cdot E)}
 +\mu \dot{\bf J}+ \mu \frac{\partial}{\partial t}\left[\dot{\theta }{\bf B}+ {\bf \nabla}\theta{\bf \times E}\right], \label{10}
\end{equation}
\begin{equation}
\nabla^2 {\bf H}-\varepsilon\mu \ddot{\bf H}= -{\bf \nabla \times J}-{\bf \nabla \times }[\dot{\theta}{\bf B}+{\bf \nabla}\theta{\bf \times E}]. \label{11}
\end{equation}
We shall  limit ourselves to a perturbative approach in which the influence from axions is small.   We do not consider the field equations for the axions explicitly.

The field equations above are complicated in the sense that they contain the second order derivatives of $\theta$. These may conveniently be removed if we consider the approximations ,with which we are working in the following,  of assuming constant axion derivatives. We  discuss the motivations for the utility of such an approximation in the section V.

 In such a way, the field equations  take the reduced forms
 \begin{equation}
\nabla^2 {\bf E}-\varepsilon\mu \ddot{\bf E}=  {\bf \nabla (\nabla \cdot E)} +\mu {\dot{ \bf J}}+  \mu [ \dot{\theta}\dot{\bf B} + {\bf \nabla}\theta{\bf \times \dot{E}}], \label{12}
\end{equation}
\begin{equation}
\nabla^2 {\bf H}-\varepsilon\mu \ddot{\bf H} = -{\bf \nabla \times J}-\left[   \dot{\theta}{\bf \nabla \times B}
 + ({\bf \nabla}\theta){\bf \nabla \cdot E}
 -({\bf \nabla}\theta \cdot {\bf \nabla})  {\bf E}      \right]. \label{13}
\end{equation}

\section{Hybrid form of Maxwell's equations. Boundary conditions}

The purpose of this section is to show that one can construct a hybrid form of Maxwell's equations from which the generalized boundary conditions at a dielectric surface follow in a very transparent way. Moreover, this form shows the close formal relationship that exists between the Axion Electrodynamics and the usual Electrodynamics for a chiral medium.

Introduce new fields ${\bf D}_\gamma$ and ${\bf H}_\gamma$ via
\begin{equation}\label{25}
{\bf D}_\gamma= \varepsilon {\bf E}+\theta{\bf B}, \quad {\bf H}_\gamma= {\bf H}-\theta {\bf E}.
\end{equation}
When written as
\begin{equation}
\left(\begin{array}{ll}
{\bf D}_\gamma \\
{\bf H}_\gamma
\end{array}
\right)=
\left( \begin{array}{ll}
\varepsilon  & \theta \\
-\theta      & 1/\mu
\end{array}
\right)
\left(\begin{array}{ll}
{\bf E} \\
{\bf B}
\end{array}
\right),
\end{equation}
it is seen how ${\bf D}_\gamma, {\bf H}_\gamma$  relate to the response tensor $H^{\mu\nu}$ and not to the original field tensor $F_{\mu\nu}$.
In terms of the new fields, the Maxwell equations get formally the same appearance  as in usual electrodynamics,
\begin{equation}
{ \bf \nabla \times H}_\gamma = {\bf J}+ \dot{\bf D}_\gamma, \quad {\bf \nabla \cdot D}_\gamma = \rho,
\end{equation}
\begin{equation}
  { \bf \nabla \times E} = -       \dot{\bf  B},  \quad {\bf \nabla \cdot B} =0,
\end{equation}
although they have now a hybrid form.
The  boundary conditions at a dielectric boundary are then immediate,
\begin{equation}
E_\perp = E_{\gamma,\perp} + E_{a,\perp} \quad \rm{ is ~continuous},
\end{equation}
\begin{equation}
H_{\gamma,\perp} \quad \rm{is ~continuous},
\end{equation}
where the symbol $\perp$ means perpendicular to the normal, thus parallel to the surface.

An important quantity in this context is the Poynting's vector, ${\bf S= E\times H}$.
Take the $z$ component $S_z$ orthogonal to a dielectric interface between a medium $1$ and a medium $2$:
\begin{equation}
S_{1z}=E_{1,\perp}H_{1,\perp}= E_{1,\perp}(H_{1,\perp}^\gamma +\theta E_{1,\perp}),
\end{equation}
\begin{equation}
S_{2z}=E_{2,\perp}H_{2,\perp}= E_{2,\perp}(H_{2,\perp}^\gamma +\theta E_{2,\perp}).
\end{equation}
As $E_\perp$ and $H_\perp^\gamma$ are continuous it follows that
\begin{equation}
S_{1z}=S_{2z},
\end{equation}
showing that the energy flow is continuous across the surface. This is as one would expect as the surface is {\it at rest}; the force acting on it is not able to do any mechanical work.

\section{Dispersion  relations in the case of infinite width}

We now consider electromagnetic waves emanating from the metal surface $z=0$ in the $+z$ direction, and consider in this section  the limit for which the width $L$ of the slab is infinite. That implies that the situation is {\it translationally invariant} and we want to evaluate the dispersion relations of the photons in our special case of Axion Electrodynamics. It means that we can introduce the wave component $k_z \equiv k$ in the $z$ direction also, and so we can make use of the standard plane wave expansion
\begin{equation}
{\bf E} \propto e^{i({\bf k\cdot x}-\omega t)}, \quad  {\bf  k}= (k_x,k_y,k). \label{expansion}
\end{equation}
We start from the reduced Maxwell equations (\ref{12}) and (\ref{13}), assume $a(x)$ to vary with space and time, but restrict ourselves to cases where the derivatives are constants. Thus, in terms of two new symbols
\begin{equation}
\alpha = \dot{\theta}, \quad {\bf \beta}= {\bf \nabla}\theta,
\end{equation}
we assume $\alpha =$ constant, ${\bf \beta} =$constant.

The motivations for such a restriction are several. This can be a good approximation when we can assume the axion field to be slowly varying on the typical distance and the timescale set up by the typical wavelenghts of the light and, obviously, the typical ones of the physical system of interest. Here we distinguish two cases and discuss the orders of magnitude with the High Energy Physics axion :
	\begin{enumerate}
\item 	$\alpha=0$: This means dealing with a static axion configuration, for example a domain wall. In such a case, our approximation is good if the distance scale $L$ of our Casimir set-up, that is also the wavelenght scale of the electromagnetic modes giving a more significant contribution to Casimir effect,  is much smaller than the typical dimension of the axion configuration, that is the inverse $m_a^{-1}$ of the axion mass in both the two cases.

If we consider $ 1 ~\mu m$ as a typical distance scale for Casimir systems (see cf. Refs~\cite{lamoreaux97,lamoreraux11,bimonte16}  ) this means an axion mass much lower than $10^{-1}~$eV and this is clearly consistent with the current bounds on the axion mass. In such a case, an order of magnitude of $\beta$ is $\sim 10^{-2} \pi m_a$ and the axionic correction to the Casimir force at zero temperature is $\sim \beta^2/L^2$ by dimensional means, so if we take the nonaxionic Casimir force $\sim \frac{\pi^2}{240 L^4}$ we expect the axion contribution to be of the same order, in SI units, for $L \gtrsim 100~\mu m \left( \frac{10^{-2}~eV}{m_a}\right) $.

This distance is at the moment too big for current Casimir set-ups (see \cite{lamoreaux97,lamoreraux11,bimonte16}  , for example Lamoreaux \cite{lamoreaux97} has measured the non thermal Casimir force between a sphere and a plate at a maximum distance of $6~\mu m$), so it could be of interest for future experimental set-ups to possibly explain anisotropies or setting experimental constraints on possible ALP domain walls in the current Universe. Indeed, in the following we perform calculations assuming $\beta$=constant and directed along $z$, so the two plates are exactly aligned with the domain wall, but this cannot be true in a real experiment. Consequently we expect also for an actual axion correction a behaviour $\propto \cos^2{\Omega}$ where $\Omega$ is the angle between the normal direction of the two plates and the one of the domain wall, so the presence of an axion domain wall could lead to anisotropic effects on the Casimir force, that could be possibly easier to detect experimentally.

\item $\beta=0$: This means dealing with a homogenous axion field, but time-dependent. In such a case an obvious axion configuration is the configuration    with a typical time scale that is $\sim m_a^{-1}$, corresponding in SI units to $\sim 10^{-3}~s~ \left(\frac{10^{-12} ~eV}{m_a}   \right)$. If we assume a time of $10^4 ~s$ to be a large timescale, this means that it could be a good approximation for $m_a<10^{-19}~eV$. It is indeed worth noticing that axions with mass $m_a \sim 10^{-23}-10^{-22}~eV$ are interesting as \emph{warm} dark matter candidates. An order of magnitude for $\alpha$ for a cosmological dark matter axion in the current Universe would be $\alpha \sim 10^{-21}~ m_a^{-1}   $, however it could be of the order of $\sim 10^{-2}\pi~ m_a^{-1}$ either in the Early Universe or inside some neutron stars . In such two last cases, the order of magnitude of the corrections to Casimir force are the same as suggested in the former point for $\alpha=0$. Differently from the former case, we would not have anisotropic effects on the Casimir force.
	\end{enumerate}
It is worth to remark the difference in the suggested axion masses in the two former points, deriving from the fact that in the first point the typical spatial dimensions of the system of interest need to be smaller than $m_a^{-1}$ \emph{as a distance scale}, while in the second one the typical time scales need to be smaller than $m_a^{-1}$ \emph{as a time scale}.

Furthermore, and not less importantly, there exist topological materials, such as Weyl semimetals,  which Chern-Simons effective interaction with photons is formally the same of Axion Electrodynamics, with an effective axion field that is with constant derivatives inside the material (see cf. \cite{anand18} for a review). In such a case, we can have for both $\alpha$ and $\beta$ an order of magnitude that can be of $10^{-5} ~eV$ or above.

We start from Eqs.~(\ref{12}) and (\ref{13}), assuming $\rho={\bf J} =0$,
\begin{equation}
\nabla^2 {\bf E}-\varepsilon\mu \ddot{\bf E} =\mu (\alpha \dot{{\bf B}}+{\bf \beta}\times \dot{{\bf E}}), \label{governingequation}
\end{equation}
\begin{equation}
\nabla^2 {\bf H}-\varepsilon\mu \ddot{\bf H} = -\alpha {\bf \nabla \times B }- {\bf \beta}{\bf \nabla \cdot E} + {\bf (\beta \cdot \nabla)E}.
\end{equation}
From the equation for $\bf E$, inserting the  plane wave expansion (\ref{expansion}), we  obtain the following set,
\begin{equation}
\left( \begin{array}{lll}
-({\bf k}^2+\varepsilon\mu\omega^2) & -i\mu(\alpha k+\beta\omega) & i\mu \alpha k_y  \\
i\mu (\alpha k+\beta \omega) & -({\bf k}^2+\varepsilon\mu\omega^2) & -i\mu\alpha k_x \\
-i\mu \alpha k_y & i\mu \alpha k_x & -({\bf k}^2+\varepsilon\mu\omega^2)
\end{array}
\right)
\left( \begin{array}{lll}
E_x \\
E_y \\
E_z
\end{array}
\right) =0, \label{E}
\end{equation}
where $\beta = d\theta/dz$ and ${\bf k}^2=k_x^2+k_y^2+k^2$.

Requiring the system determinant to be zero, one obtains the dispersion relations. There are two dispersive branchesequation. The first is a "normal" one,
\begin{equation}
\varepsilon\mu \omega^2 = {\bf k}^2,
\end{equation}
corresponding to waves completely independent of the axions. The second branch is
\begin{equation}
\varepsilon\mu \omega^2= {\bf k}^2 \pm \sqrt{\alpha^2k_\perp^2 + (\alpha k+\beta \omega)^2}, \quad k_\perp^2=k_x^2+k_y^2,
\end{equation}
showing the splitting of this branch into two modes, equally separated from the normal mode on both sides. This sort of splitting has been encountered before under various circumstances; cf., for instance, Refs.~\cite{sikivie03,mcdonald20,brevik21a}.

It should be noticed that there is so far no restriction on the magnitudes of $\alpha$ and $\beta$. For practical purposes, it will be convenient to discuss the cases $\alpha=0$ and $\beta=0$ separately.

\subsection{The case $\alpha=0$}

This is the situation often discussed in connection with topological insulators. The dispersion relations for the nontrivial branch become simple,
\begin{equation}
\varepsilon\mu \omega^2= {\bf k}^2 \pm \mu \beta \omega. \label{alpha}
\end{equation}
The following property of this expression ought to be noticed, as it relates to the ordinary electrodynamics in a nondissipative dispersive medium. Assume for example that the medium is nonmagnetic, $\mu=1$, and introduce an effective permittivity $\varepsilon_{\rm eff}$ via
\begin{equation}
{\bf k}^2 = \varepsilon_{\rm eff}(\omega) \omega^2. \label{dispersion}
\end{equation}
Then consider the expression for the electromagnetic energy density in this kind of  medium (in complex representation),
\begin{equation}\label{35}
W_{\rm disp}= \frac{1}{4}\left[\frac{d(\varepsilon_{\rm eff}\omega)}{d\omega}|{\bf E}|^2+|{\bf H}|^2\right].
\end{equation}
From the dispersion relation in this case it follows  that
\begin{equation}
\frac{d(\varepsilon_{\rm eff}\omega)}{d\omega} = \varepsilon,
\end{equation}
showing that  the dispersive correction disappears. The energy density is the same as if dispersion were absent.
This restriction to a nonmagnetic medium is actually a  nontrivial point .We follow here the conventional formalism, as presented, for instance, by L. D. Landau and E. M. Lifshitz, Electrodynamics of Continuous Media \cite{landau}, for the electromagnetic theory for slightly dispersive media. In principle, one might include the magnetic dispersion also, so as to get two additive terms in Eq.(\ref{35}).  Such a description would however require that {\it both}  the permittivity $\varepsilon$, and the permeability $\mu$, were expressed in a dispersive form. That is not the case here, as all the dispersion available is in the form of the squared refractive index $n^2 = \varepsilon \mu$ in the dispersive relation (\ref{alpha}). This combined quantity is not equivalent to $\varepsilon$ and $\mu$ separately.

\subsection{The case $\beta =0$}

The axion-influenced dispersion relation is now
\begin{equation}
\varepsilon\mu\omega^2 = {\bf k}^2 \pm \mu\alpha |{\bf k}|. \label{alfabet}
\end{equation}
We may also in this case introduce the effective permittivity $\varepsilon_{\rm eff}(\omega)$ for a nonmagnetic medium, as in Eq.~({\ref{dispersion}).  In this case it is actually  more convenient to give the expression for the effective refractive index, $n_{\rm eff}=\sqrt{\varepsilon_{\rm eff}}$.  From Eq.~(\ref{alfabet}) we get
\begin{equation}
n_{\rm eff}(\omega)= \sqrt{\varepsilon +\frac{\alpha^2}{4\omega^2}} \pm \frac{\alpha}{2\omega}.
\end{equation}
In this expression the permeability $\mu$ can easily be included also.

\bigskip

We will hereafter make use of the fact that the influence from the axions is in practice weak. We introduce the nondimensional parameter
\begin{equation}
\xi= \frac{\alpha k+\beta \omega}{k^2}, \label{ksi}
\end{equation}
(recall that $k=k_z$), and assume that $\xi \ll 1$, a reasonable assumption for an axion as a main component of dark matter, as we discuss in the following. It is indeed of interest to give some numerical estimates. We may have $k \sim \omega$ (for a photon propagating along z-direction) and associate the indicative value $m_a \sim \omega=10^{-5}~$eV and get $k=10^{-5}~$eV$\, \left(\frac{m_a}{10^{-5} \, eV}\right)$.

If such a case, the condition  $\xi \ll 1$ is roughly $\frac{\beta+\alpha }{k} \ll 1$, which leads to %$\beta+\alpha \ll 10^{-5}~$eV$\,\left(\frac{m_a}{10^{-5} \, eV}\right)$. %We may express this restriction in physical units, using the conversion factor 1~eV$= 5\times 10^4~$cm$^{-1}$,
%\begin{equation}\label{betacond}
%\beta+\alpha \ll 5\times 10^{-1}~{\rm cm}^{-1} \left(\frac{m_a}{10^{-5} \, eV}\right).
%\end{equation}
\begin{equation}\label{betacond}
\beta+\alpha \ll 10^{-5}~{\rm eV} \,\left(\frac{m_a}{10^{-5} \, eV}\right).
\end{equation}
We can observe that for a galactic plane wave axion field $\theta(x)=\theta_0 \Re{e^{i (\omega_a t-\vec{k}_a \cdot \vec{r})}}$ we have $\theta_0 \simeq 6.12 \times 10^{-19} \left( \frac{\rho}{0.45 \,\, GeV/cm^3} \right)^{1/2}$ where $\rho$ is the local axion dark matter energy density in our region of the galaxy, on which there are some uncertainties \cite{eggemeier23}.

Indeed, a value of $\rho=1.2 \times 10^{-6} \, GeV/cm^3$, corresponding to the average energy density of dark matter at cosmological scales \cite{bauman2022}, would not correspond either to the average dark matter density in the solar neighborhood inferred by astronomical means or to the typical dark matter density sampled by an experiment. As recently and widely discussed in Ref.~\cite{eggemeier23},these last two densities can be different if the local axion structure is formed by several minivoids and not with an homogenous structure, in particular a typical experiment would sample at a given instant $\sim 10 \%$ of the average density inferred astronomically, whose value is accordingly to most recent analyses in the range of $0.2-0.7 \,\, GeV/cm^3$ \cite{de Salas_2021}. Usually, it is adopted the value $0.45 \,\, GeV/cm^3$ in the axion literature \cite{liu22,eggemeier23}.

$\beta \sim \theta_0 \lambda_{dB}^{-1} \sim 10^{-29} \,$eV$\,  \left(\frac{m_a}{10^{-5} \, eV}\right) $, where $\lambda_{dB}= \frac{2 \pi}{m_a v}$ is the de Broglie wavelength. We took $v \sim 10^{-3}$, while $\alpha \sim \theta_0 m_a \sim 10^{-26} \,$eV$\,  \left(\frac{m_a}{10^{-5} \, eV}\right) $. Consequently, the condition (\ref{betacond}) would be trivially satisfied in such a case (and also with much bigger values of $\theta_0$).%ADJUST

 For a cosmological axion domain wall we may take $\alpha \approx 0$, $\beta \sim \theta m_a \sim 10^{-7} \,$eV$\,  \left(\frac{m_a}{10^{-5} \, eV}\right) $, where $\theta \sim g_{\gamma } \alpha \sim 10^{-2}$, so the condition (\ref{betacond}) is also satisfied.
 %If $\beta =0$, the analogous condition is $\alpha/k \ll 1$, which means $\alpha \ll 10^{-8}~$eV. Again going over to physical units, now using the conversion factor in the form $1~$eV $=1.5\times 10^{15}~$s$^{-1}$, we get
%\begin{equation}
%\alpha \ll 1.5\times 10^7~{\rm s}^{-1}.
%\end{equation}

\section{The case of finite width}

If the slab width $L$ is finite, the system is no longer translationally  invariant in the $z$ direction. This is an important difference from the case discussed in the previous section, as we can no longer make use of the plane wave expansion (\ref{expansion}) in full. What can be taken over to the present case are the transverse vector components $k_x$ and $k_y$, but the longitudinal component $k_z$ not. We will now solve the governing equations for the electric field perturbatively, to the first order in the smallness parameter $\xi$ defined in Eq.~(\ref{ksi}), and investigate how the basic modes develop directly in configuration space, in the $z$ direction.

We assume the following form for the fields,
\begin{equation}
{\bf E}({\bf x},t)= {\bf E}(z)e^{i\Phi}, \quad \Phi= {\bf k_\perp \cdot x_\perp}-\omega t,
\end{equation}
 and start from the governing equation (\ref{governingequation}) for $\bf E$. The right hand side is small, and we can therefore use on this side the expressions for the TE modes given earlier in Eq.~(\ref{TEmode}). We set the normalization constant $N$ equal to 1 for simplicity. It is convenient to keep the symbol $k=k_z$ in these expressions, restricted as before by the condition $k=\pi p/L$, although this symbol serves only as a calculational tool in the approximate calculation. To reemphasize, this does not mean that we assume translational invariance.

 We define $\lambda^2$ as
 \begin{equation}
 \lambda^2= \varepsilon\mu\omega^2-k_\perp^2, \label{kappa}
 \end{equation}
and write out all three component equations,
\begin{equation}
E_x''(z)+\lambda^2 E_x(z)= i\mu k^2\xi E_y(z)-i\mu\alpha k_y E_z(z), \label{modified}
\end{equation}
\begin{equation}
E_y''(z)+\lambda^2 E_y(z) = -i\mu k^2\xi E_x(z) +i\mu \alpha k_x E_z(z),
\end{equation}
\begin{equation}
E_z''(z)+\lambda^2 E_z(z) =i\mu \alpha \omega k_\perp^2\sin kz.
\end{equation}
These  modified  equations correspond to the earlier equations (\ref{E}) in the translationally invariant case.

We solve the equation for $E_x(z)$ as an inhomogeneous differential equation (cf., for instance, p 530 in Ref.~\cite{morse53}),  observing that the two basic solutions for the homogeneous equation can be chosen as $\psi_1= \sin \lambda z$ and $\psi_2= \cos \lambda z$, with Wronskian $\psi_1 \psi_2'-\psi_2\psi_1'=-\lambda$.  For the fields on the right hand side of Eq.~(\ref{modified}) we insert the TE expressions from Eq.~(\ref{TEmode}). We write the solution $E_x(z)$ as a sum of two terms,
\begin{equation}
E_x(z)= E_x^{(1)}(z)+  E_x^{(2)}(z),
\end{equation}
where $E_x^{(1)}$ and $E_x^{(2)}$ refer respectively to $\psi_1$ and $\psi_2$. Some calculation leads to the expressions
\begin{equation}
E_x^{(1)}(z)= C_1\sin \lambda z -\frac{i}{2}(\mu \omega N)kk_x\xi\left[ \frac{1-\cos (k-\lambda)z}{k-\lambda}+\frac{1-\cos (k+\lambda)z}{k+\lambda}\right]\sin \lambda z,
\end{equation}
\begin{equation}
E_x^{(2)}(z)=   C_2\cos \lambda z   +  \frac{i}{2}(\mu \omega N)kk_x \xi \left[ \frac{\sin(k-\lambda) z}{k-\lambda}-
\frac{\sin(k+\lambda)z}{k+\lambda}\right] \cos \lambda z,
\end{equation}
showing how the axions modify this field component to order $\xi$; cf. Eq.~(\ref{ksi}). $C_1$ and $C_2$ are constants.
  Since the difference between $\lambda$ and $k$ is small, we have replaced $\lambda$ with $k$ in the noncritical nontrigonometric terms. The expressions show that  to  first order we can make the same replacement in the trigonometric
	terms too. Requiring the total field component $E_x(z)$ to be zero at $z=0$ and $z=L$ we find that $C_1$ is undetermined, while $C_2=0$.  We can thus set $C_1= N\omega k_y$ to agree with the zeroth order expression. Altogether,  
\begin{equation}
E_x^{(1)}(z) = N\omega k_y\left[ 1- \frac{i\mu}{4}\frac{k_x}{k_y}\xi (1-\cos 2\lambda z)\right] \sin \lambda z,
\end{equation}
\begin{equation}
E_x^{(2)}(z)= \frac{i}{2}(\mu \omega N)\lambda k_x \xi \left[z-\frac{\sin 2\lambda z}{2\lambda}\right] \cos \lambda z.
\end{equation}
The imaginary terms signify a rotation of the transverse field ${\bf E}_\perp$ in the $xy$ plane. Of main interest is the rotation angle proportional to $z$,  as it is similar to the Faraday effect as well as to chiral electrodynamics. We will therefore focus on his term, and write the full component $E_x$ in the form
\begin{equation}\label{Ex}
E_x(z)= N\omega k_y[\sin \lambda z + i\gamma_x(z)\cos \lambda z].
\end{equation}

However, in order to evaluate the rotation of the optical angle we need to consider that, analogously to $E_x$ in Eq.~(\ref{Ex}), we can get the following expression for $E_y$:
 \begin{equation}\label{Ey}
 	E_y(z)= -N\omega k_x[\sin \lambda z - i\gamma_y(z)\cos \lambda z],
 \end{equation}
 where
 \begin{equation}
 	\phi_y(z)= \frac{1}{2}(\mu \lambda z)\frac{k_y}{k_x}\xi.
 \end{equation}
We now observe that we can write the usual fields $E_{\pm}(z)=E_{x}(z) \pm i E_{y}(z)$ can be written from equations (\ref{Ex},\ref{Ey}) as:
\begin{equation}\label{solpm}
E_{\pm}(z)=N \omega (k_y \mp i k_x) \left[ \sin{\lambda z} \mp \frac{1}{2}\mu (\alpha k+\beta \omega )z    \right].	
\end{equation}
In order to grasp the physical meaning of this expression we can observe that, since we work out the electric and magnetic fields up to the first order in $g_{a \gamma \gamma}$ and for TE mode we have $E_z=0$ at order zero, our results for $E_x$ and $E_y$ is equivalent to get the solution up to the first order of the equations:
\begin{equation}
	E_x''(z)+\lambda^2E_x(z)= i\mu \lambda^2\xi E_y(z), \label{modified1}
\end{equation}
\begin{equation}
	E_y''(z)+\lambda^2E_y(z) = -i\mu \lambda^2\xi E_x(z),
\end{equation}
that can be rewritten in terms of $E_{\pm}$ fields as
\begin{equation}
	E_{\pm}''(z)+[\lambda^2 \mp \mu (\alpha \lambda+\beta \omega) ]E_{\pm}(z)=0,  \label{modified2}
\end{equation}
whose general solution is
\begin{equation}\label{genpm}
	E_{\pm}(z)=A e^{i \sqrt{\lambda^2 \mp \mu (\alpha \lambda+\beta \omega)} z}+B  e^{-i \sqrt{\lambda^2 \mp \mu (\alpha \lambda+\beta \omega)} z}.
\end{equation}
If we employ the boundary conditions $E_x(z=0,L)=E_y(z=0,L)=0$ and our assumption of $\chi \ll 1$ (leading e.g. to $\sqrt{\lambda^2 \mp \mu (\alpha \lambda+\beta \omega)}  \sim \lambda \mp\frac{1}{2}  \mu \frac{\alpha \lambda+\beta \omega}{\lambda} $), then we get the same solution (\ref{solpm}) .
Now the physical meaning of the solution (\ref{solpm}) is clear thanks to the expression (\ref{genpm}): the phase velocities of left and right circularly-polarized waves are respectively different (see Ref.~\cite{jackson}), so the optical angle rotates from $z=0$ to $z$ of the angle
\begin{equation}
\phi(z)=\frac{1}{2}  \frac{\mu}{n} \frac{\alpha \lambda+\beta \omega}{\lambda}z=\frac{1}{2}  \sqrt{\frac{\mu}{\varepsilon}} \frac{\alpha \lambda+\beta \omega}{\lambda}z.	
\end{equation}
 This rotation of the optical angles consequently results on a gradual transition of the TM mode into a TE mode, and similarly in the reverse direction TE $\rightarrow$ TM.
The value of $\phi$ at $z=L$ is then seen to be
\begin{equation}
\phi(L)= \frac{1}{2}  \sqrt{\frac{\mu}{\varepsilon}}\frac{\alpha \lambda+\beta \omega}{\lambda}L. \label{fi}
\end{equation}
This result is consistent with similar expressions obtained previously by Refs.~\cite{carroll90,sikivie03,favitta23}.

\section{Casimir effect}

We assume the same system as in the previous section: the regions $z<L$ and $z>0$ are perfectly conducting, and the intermediate region $0<z<L$ filled with a uniform  dielectric with material constants $\varepsilon$ and $\mu$. The axion field is also assumed to fill the intermediate region. This field may in principle vary both in space and time, but we assume as before that the parameter $\xi$ is small; cf. the definition (\ref{ksi}). There is no external magnetic field.

We intend to calculate the Casimir free energy $F$ between the plates per unit surface area, and begin with the known expression from ordinary  (axion-free) electrodynamics at temperature $T$,
\begin{equation}
F= \frac{1}{\pi\beta}{\sum_{m=0}^\infty}¨^ \prime \int_{n\zeta_m}^\infty \kappa d\kappa \ln (1-e^{-2\kappa L}). \label{energy}
\end{equation}
Here  $\zeta_m=2\pi m/\beta_T$ with $\beta_T =1/T$ is the Matsubara frequency, and $\kappa$ is defined by $\kappa^2=k_\perp^2+n^2\zeta^2$ with $n^2=\varepsilon\mu$. Note that $\kappa$ is defined here in a conventional way (cf., for instance, Refs.~\cite{milton01,brevik08}). The quantity $\lambda$ defined in Eq.~(\ref{kappa}) is different, although physically related.

To put our approach into a  wider perspective, we will first recapitulate briefly two related situations:

\noindent {\bf 1.~} First, consider the purely electromagnetic case with a chiral medium between the plates, when there is also a strong external magnetic field ${\bf B}_0$ in the $z$ direction. Both modes, TE and TM, will rotate between the plates. One may read this problem as an interaction between harmonic oscillators in the two plates. The result is that there occurs a slow rotation of the polarization plane, proportional to $z$ as well as to $B_0$,  as the wave propagates through the medium. There occurs a gradual transition of the TM mode into a TE mode, and similarly in the reverse direction TE $\rightarrow$ TM.

Let $\phi$ denote the rotation angle at $z=L$. Of physical interest is the rotation matrix
\begin{equation}
{\bf A} = \left( \begin{array}{ll}
\cos \phi & \sin \phi \\
-\sin \phi & \cos \phi
\end{array}\right).
\end{equation}
When the wave travels back, the important point is whether the rotation occurs in the reverse direction, thus $\phi =0$ in total, or if the rotation continues in the same direction, so that $\phi \rightarrow 2\phi$ in total. Only the last case leads to physical effects. We therefore have to do with the square of the matrix above,
\begin{equation}
{\bf A^2} = \left( \begin{array}{ll}
\cos 2\phi & \sin 2\phi \\
-\sin 2\phi & \cos 2\phi
\end{array}\right).
\end{equation}
This transformation  matrix, when inserted into the Casimir energy formula, was in Ref.~\cite{hoye20} found to lead to the same answer as derived earlier in Ref.~\cite{jiang19}, in a more compact way. It should also be  mentioned here that the Faraday effect in a optically active material, in the presence of a longitudinal magnetic field, is closely related. There is then a rotation of the polarization plane  proportional to $z$, $\phi(z) = {\cal V}B_0 z$, where the material constant $\cal V$ is called the Verdet constant.

\noindent {\bf 2.~} Another known case of considerable interest is the so-called Boyer problem \cite{boyer74}, where one of the metal plates is replaced by an ideal "magnetic" plate.  This case corresponds to the rotation angle $\phi = 90^o$, and leads actually to a repulsion between the two plates. A further discussion of the Boyer problem can be found, for instance, in Ref.~\cite{hoye18}. 

\bigskip
In all of these cases the rotation of the optical angle gives a gradual transition of the TM mode into a TE mode, and similarly in the reverse direction TE $\rightarrow$ TM, when an electromagnetic wave propagates.

We return to the axion problem, following the same method as anticipated above. We first observe that the logarithmic factor in the energy expression (\ref{energy}) can be written as a trace,
\begin{equation}
2\ln (1-e^{-2\kappa L}) = {\rm Tr} [\ln ({\bf I}-e^{-2\kappa L}{\bf I})],
\end{equation}
where $\bf I$ is the unit matrix in two dimensions. We now replace $\bf I$ with the round-trip matrix $\bf A^2$ in the interaction term, containing the exponential. It gives  rise to the effective substitution
 \begin{equation}
 2\ln (1-e^{-2\kappa L}) \rightarrow {\rm Tr}[\ln ({\bf I}-e^{-2\kappa L}{ \bf A}^2)] = \ln[\det ({\bf I}-e^{-2\kappa L}{ \bf A}^2)].
 \end{equation}
Here, $\phi$ means the axion rotation angle $\phi(L)$ as given in Eq.~(\ref{fi}). As the determinant is
\begin{equation}
\det({\bf I}-e^{-2\kappa L}{ \bf A}^2) = 1+e^{-4\kappa L}-2e^{-2\kappa L}\cos 2\phi,
\end{equation}
we obtain from Eq.~(\ref{energy}) the following expression for the Casimir free energy,
\begin{equation}
F= \frac{1}{2\pi \beta_T}{\sum_{m=0}^\infty}¨^ \prime \int_{n\zeta_m}^\infty \kappa d\kappa \ln(1+e^{-4\kappa L} -2e^{-2\kappa L}\cos 2\phi), \label{freeenergy}
\end{equation}
or more explicitly
\begin{equation}
F= \frac{1}{2\pi \beta_T}{\sum_{m=0}^\infty}¨^ \prime \int_{n\zeta_m}^\infty \kappa d\kappa \ln\left[1+e^{-4\kappa L} -2e^{-2\kappa L}\cos\left(\sqrt{\frac{\mu}{\varepsilon}}\frac{\alpha \kappa+\beta \zeta_m}{\kappa}L\right) \right]. \label{freeenergy1}	
\end{equation}
We note that with $\beta=0$ the phase $2\phi=\sqrt{\frac{\mu}{\varepsilon}}\alpha L$ is not dependent on $\kappa$ and $\zeta_m$.

The only place where the presence of axions shows up, is in the phase $\phi$. The formula combines in a unified fashion the space and the time-varying axion field.
The expression (\ref{freeenergy}) is formally the same as for a chiral medium, and has a wide applicability.
For instance, for ideal metal plates in the nonaxion case ($\phi=0$), we have
\begin{equation}
F_{\rm  metal} = \frac{1}{\pi\beta_T}{\sum_{m=0}^\infty}¨^ \prime \int_{n\zeta_m}^\infty \kappa d\kappa \ln (1-e^{-2\kappa L}),
\end{equation}
whereas  in the repulsive Boyer case ($\phi = 90^o)$,
\begin{equation}
F_{\rm Boyer}= \frac{1}{\pi\beta_T}{\sum_{m=0}^\infty}¨^ \prime \int_{n\zeta_m}^\infty \kappa d\kappa \ln(1+e^{-2\kappa L}).
\end{equation}
Finally, at zero temperature the free energy $F$ reduces to the thermodynamic energy $E$.  Making use of the relationship
\begin{equation}
\frac{1}{\beta_T}{\sum_{m=0}^\infty}^ \prime \rightarrow \frac{1}{2\pi}\int_0^\infty d\zeta,
\end{equation}
we then obtain the zero temperature variant of Eq.~(\ref{freeenergy}),
\begin{equation}
E_{T=0} = \frac{1}{(2\pi)^2}\int_0^\infty d\zeta \int_{n\zeta}^\infty \kappa d\kappa \ln (1+e^{-4\kappa L}- 2e^{-2\kappa L}\cos 2\phi). \label{nullpunktsenergi}
\end{equation}
As before, $\kappa^2=k_\perp^2+n^2\zeta^2$, but now with $\zeta$ as a continuous variable.

It is noteworthy that for small rotation angles $\phi$, the corrections from axions occur to the order $\phi^2$. We may express this in a more explicit fashion by rewriting Eq.~(\ref{freeenergy}) as
\begin{equation}
F= F_{\rm metal} -\frac{4}{\pi \beta} {\sum_{m=0}^\infty}^\prime  \int_{n\zeta_m}^\infty  \kappa d\kappa \frac{e^{-2\kappa L}}{(1-e^{-2\kappa L})^2} \phi^2.\label{freenergy2ndorder}
\end{equation}
From the former results we can get some interesting results for particular cases of interest.
\subsection{The case $\beta=0$}
In this case the expression (\ref{freeenergy1}) becomes

\begin{equation}
	F= \frac{1}{2\pi \beta_T}{\sum_{m=0}^\infty}¨^ \prime \int_{n\zeta_m}^\infty \kappa d\kappa \ln(1+e^{-4\kappa L} -2e^{-2\kappa L}\cos\left(\sqrt{\frac{\mu}{\varepsilon}} \alpha L\right)), \label{freeenergybeta0T}	
\end{equation}
and the expression (\ref{freenergy2ndorder}) at the second order becomes

\begin{equation}
	F= F_{\rm metal} -\frac{1}{2\pi \beta_T} \frac{\mu}{\varepsilon}( \alpha L)^2{\sum_{m=0}^\infty}^\prime  \int_{n\zeta_m}^\infty  \kappa d\kappa \frac{e^{-2\kappa L}}{(1-e^{-2\kappa L})^2} \,.\label{freenergy2ndorderbeta0T}
\end{equation}
\subsubsection{Limit for $T=0$}
The last expression (\ref{freenergy2ndorderbeta0T}) can be evaluated straightforward in the case $T=0$, using that the double integral can be evaluated using
\begin{equation}
	\int_0^{+ \infty} d\zeta'\int_{\zeta'}^{+\infty} \kappa' d\kappa' \frac{e^{-2 \kappa'}}{(1-e^{-2 \kappa'})^2}=\frac{\zeta(2)}{2}=\frac{\pi^2}{24},
\end{equation}
so we get
\begin{equation}\label{alphaE0}
	E_{T=0}-E_{T=0,\rm metal}=-\frac{1}{48} \frac{\mu^{1/2}}{\varepsilon^{3/2}}\alpha^2 \frac{1}{L},
\end{equation}
that, if taken with $\mu=\varepsilon=1$, gives a result similar to the one in Ref.~\cite{favitta23}.
\subsubsection{Limit for $T \rightarrow +\infty$}
\begin{figure}[h]
	\includegraphics[scale=0.2,width=0.5\paperwidth]{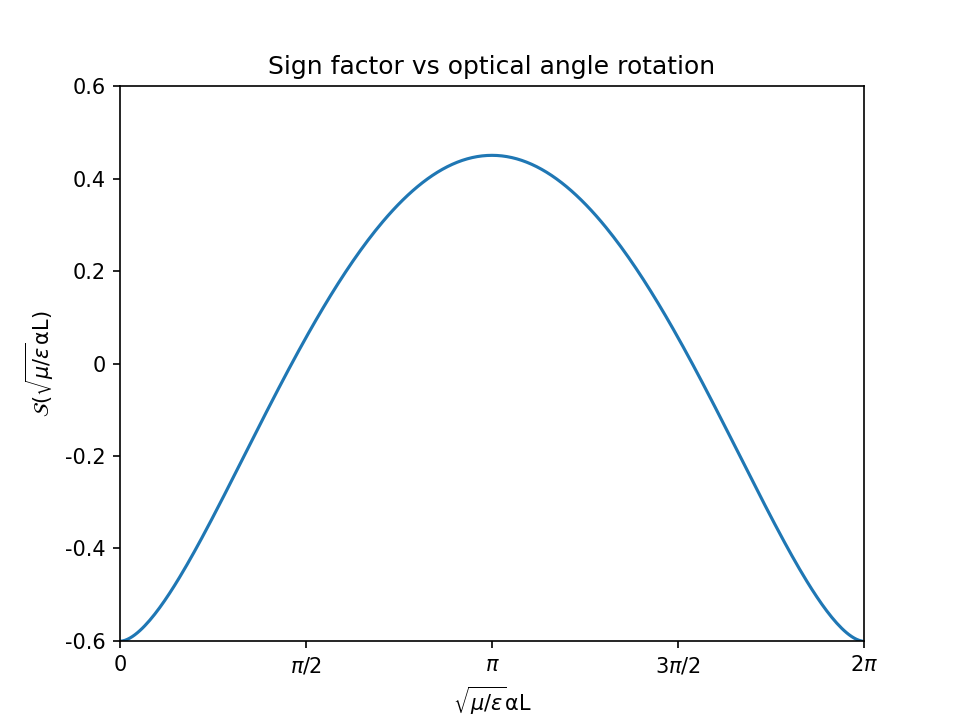}
	\caption{Plot of the sign factor as a function of $2 \phi=\sqrt{\frac{\mu}{\varepsilon}} \alpha L$}\label{fig:1}
\end{figure}
	
\begin{figure}
	\includegraphics[scale=0.2,width=0.5\paperwidth]{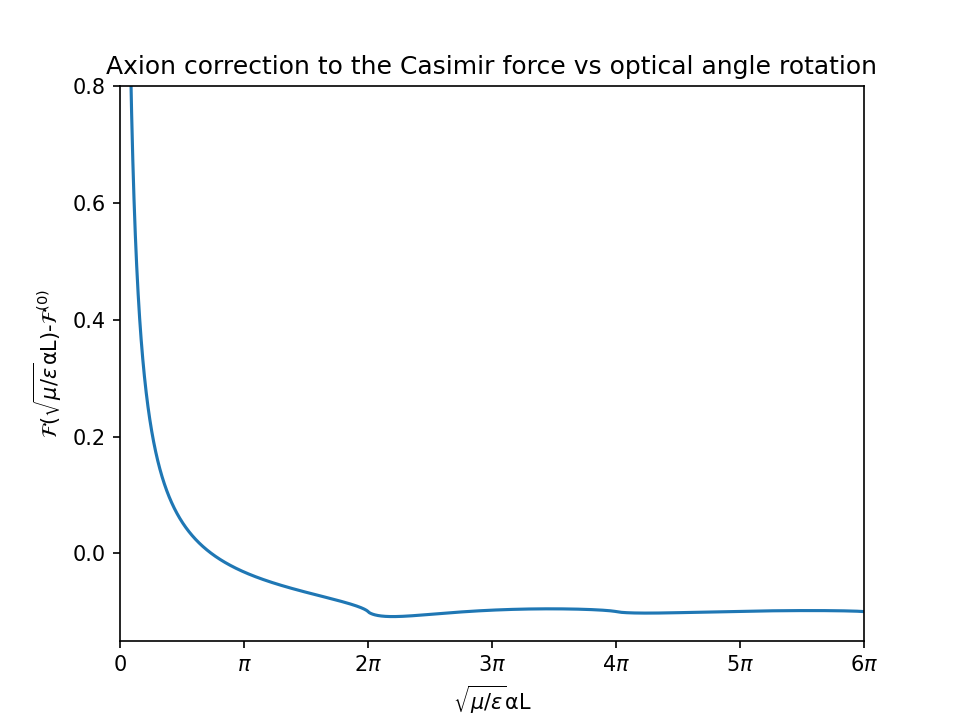}
	\caption{Plot of the ratio $\mathcal{F}\left(\sqrt{\frac{\mu}{\varepsilon}}\alpha L\right)=4 \pi \beta \frac{f^{T \rightarrow +\infty}(L,\alpha)-f^{T \rightarrow +\infty}(L,\alpha=0)}{ (\frac{\mu}{\varepsilon})^{3/2} \alpha^3}$ as a function of $2 \phi=\sqrt{\frac{\mu}{\varepsilon}} \alpha L$.For $\phi \ll \pi/2$ the axion correction is repulsive, so very differently from the case $T=0$ where it is attractive. However for $2 \phi \sim \pi$ the axion term becomes attractive.  }\label{fig:2}
	\includegraphics[scale=0.2,width=0.5\paperwidth]{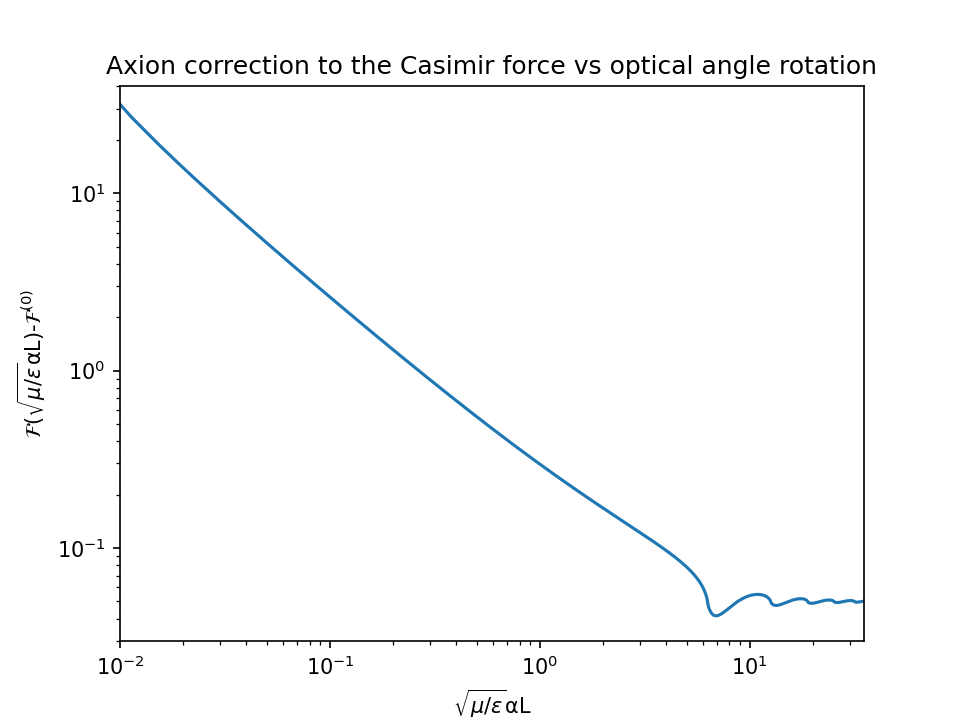}
	\caption{The same plot of Figure (\ref{fig:2}) in a log-log graph and where we have substracted the minimum to have only positive values in the y-axis.It highlights the behaviour of the axion correction to Casimir force for $\phi \ll \pi/2$, that is $\sim 1/L$ differently from the $T=0$ case. Significant deviation from such a behaviour is for $\phi >\pi/2$ as also shown in the same Figure (\ref{fig:2})} \label{fig:3}
\end{figure}
In such a case, as done in usual Casimir calculation, we get this limit by only considering the first term $m=0$ in the series and can evaluate it exactly from expression (\ref{freeenergybeta0T}):
\begin{equation}\label{ftinfinity}
	F^{T \rightarrow +\infty}=\frac{T}{4 \pi L^2} \int_0^{+ \infty} x\, dx \ln\left[1+e^{-4x}-2e^{-2x} \cos{\sqrt{\frac{\mu}{\varepsilon}} \alpha L}\right] \vcentcolon = \frac{T}{4 \pi L^2} \mathcal{S}\left(\sqrt{\frac{\mu}{\varepsilon}} \alpha L \right).
\end{equation}
We call the function $\mathcal{S}$ as a sign factor for the sake of simplicity and can be evaluated numerically.
We show its plot in Figure (\ref{fig:1}).
In order to clarify if such a behaviour is significant for the properties of the Casimir force, if it is repulsive or attractive, we plot in Figure (\ref{fig:2}) the behaviour of the Casimir force, calculated as:
\begin{equation}
	f^{T \rightarrow +\infty}(L)=-\frac{\partial 	F^{T \rightarrow +\infty}}{\partial L}= \frac{2}{L}		F^{T \rightarrow +\infty}-\frac{T}{4 \pi L^2}  \mathcal{S}'\left(\sqrt{\frac{\mu}{\varepsilon}} \alpha L \right),
\end{equation}
and we substract to it the notourios expression of the Casimir force in the same temperature limit in the usual electrodynamics:
\begin{equation}
	f^{T \rightarrow +\infty}(L,\alpha=0)=-T \frac{\zeta(3)}{8 \pi L^3}.
\end{equation}

We observe how for $\phi \ll \pi/2$ the axion correction goes as $\sim 1/L$ (as shown better in Figure (\ref{fig:3}) ) and it is repulsive, so very differently from the case $T=0$ where it goes as $1/L^2$ and it is attractive. However for $2 \phi \sim \pi$ the axion term becomes attractive.
It is worth to notice from Figure (\ref{fig:1}) that the sign factor has its absolute maximum at $\frac{\mu}{\varepsilon} \alpha L=\pi$ and this value corresponds roughly to the threshold between repulsive and attractive regime, as visible in the figures (\ref{fig:2}) and (\ref{fig:3}). This value corresponds to a value of $\alpha$ that is roughly equal to the inverse distance $L^{-1}$ and corresponds to the physical condition of maximum reflection of photons due to the presence of the axion domain wall (see the system treated widely in Ref.~\cite{favitta23}, composed by a single domain wall and no slabs, where it is shown that the axion domain wall has an analogous maximum reflectance. The correspondence between the two holds with $\omega \leftrightarrow 1/L$).

Another interesting property of the expression (\ref{ftinfinity}), that is present in the general expression (\ref{freeenergy1}),  is that, apart of a factor $L^{-2}$, we deal with an integral dependent on the double of the optical rotation $\frac{\mu}{\varepsilon} \alpha L$ and, in particular, such integral is periodic in the same angle. This leads to the observable wiggles in the Figures (\ref{fig:2}) and (\ref{fig:3}) at $2 \phi=2 n \pi$, where $n=1,2,...$.
      
\subsection{The case $\alpha=0$}
In this case the expression (\ref{freeenergy1}) becomes

\begin{equation}
	F= \frac{1}{2\pi \beta_T}{\sum_{m=0}^\infty}¨^ \prime \int_{n\zeta_m}^\infty \kappa d\kappa \ln\left(1+e^{-4\kappa L} -2e^{-2\kappa L}\cos\left(\sqrt{\frac{\mu}{\varepsilon}} \beta \frac{\zeta_m}{\kappa} L\right)\right),	
\end{equation}
and the expression (\ref{freenergy2ndorder}) at the second order becomes

\begin{equation}
	F= F_{\rm metal} -\frac{1}{2\pi \beta_T} \frac{\mu}{\varepsilon}{\sum_{m=0}^\infty}^\prime  \int_{n\zeta_m}^\infty  \kappa d\kappa \frac{e^{-2\kappa L}}{(1-e^{-2\kappa L})^2} ( \beta L)^2 \frac{\zeta_m^2}{\kappa^2}L^2.
\end{equation}
\subsubsection{Limit for $T \rightarrow +\infty$}
This is easier because, as in the case of $\beta=0$, we develop such limit by only taking $m=0$ and we get simply the nonaxionic expression:
\begin{equation}
	F=\frac{1}{2\pi \beta_T} \int_{0}^\infty \kappa d\kappa \ln\left(1-e^{-2\kappa L} \right),
\end{equation}
whose result is the notorious high temperature limit \cite{milton01}:
\begin{equation}
	F=-\frac{\zeta(3)}{8 \pi L^2},
\end{equation}
meaning that the axion correction is suppressed in the high temperature limit.

\subsubsection{Limit for $T=0$}
In such a case, the expression at the second order becomes:
\begin{equation}
	F= F_{\rm metal} -\frac{1}{(2 \pi)^2} \frac{\mu}{\varepsilon} \int_0^{+\infty} d\zeta  \int_{n\zeta}^\infty  \kappa d\kappa \frac{e^{-2\kappa L}}{(1-e^{-2\kappa L})^2} \beta^2 \frac{\zeta^2}{\kappa^2}L^2 .	
\end{equation}
This can be evaluated by a change of variables and using the numerical result of the integral:
\begin{equation}
\iota=\int_0^{+\infty} ds \int_s^{+\infty} dk \,\frac{e^{-2k}}{(1-e^{-2k})^2} \frac{s^2}{k}=0.137078,
\end{equation}
from which get, similarly to the case $\beta=0$, the attractive term:
\begin{equation}
F-F_{\rm metal}=-\iota \frac{1}{(2 \pi)^2} \frac{\mu^{1/2}}{\varepsilon^{3/2}} \frac{\beta^2}{L},
\end{equation}
whose behaviour with the distance $L$ is the same of Eq.~(\ref{alphaE0}).
\section{Comparison with earlier works}
Concerning the  formal relation which  exists between the axion electrodynamics and the usual electrodynamics for a chiral medium expressed in the part starting from  Eq.~(\ref{25}) and further on, we would like to mention the following.

%The fact that in the chiral case considered in \cite{brevik20} there appears a strong magnetic field, while in the axion case with the formal analogy mentioned in the present work such a field  does  not appear is an interesting observation. The only possible interpretation we could think of is that in the chiral case in \cite{brevik20}, the chirality is made by a  global  external field, while in the axion electrodynamics the formal analogy is for each of the axions in the medium only as a local analogy, and in total they do not add to produce a total magnetic field.
The analogy is surely related in both cases on having a polarization proportional to the total magnetic field $\vec{B}$ and a magnetization proportional to the total electric field $\vec{E}$, and, when axion derivatives are constants, leading in both cases to a rotation of polarization plane, as formerly discussed in Refs.~\cite{carroll90,sikivie03,favitta23}.

To put our methods into some perspective, it is useful  to compare  them with those used recently by other investigators.

\noindent 1.  As regards  the Casimir energy, we find it natural to compare with  the paper of Fukushima {\it et al.} \cite{fukushima19}. This paper relates to the $T=0$ case, as well as to a vacuum environment, $\varepsilon = \mu = 1$. An important difference from our approach is that they make use of the wave vector expansion for all values of $\bf k$, including
$k_z$, for all widths $L$ of the slab. That is, they follow the same approach as we did in Sec.~V, thus ignoring the lack of translational invariance for finite width. In this way, it becomes simple to calculate the Casimir energy, namely as a sum over discrete modes (their equation 21),
\begin{equation}
E_{T=0}= \sum_\pm \sum_{m=0}^\infty \int \frac {dk_xdk_y}{(2\pi)^2}\,\frac{\omega_\pm}{2}.
\end{equation}
 This is different from our logarithmic expression (\ref{nullpunktsenergi}), where $\zeta$ was a continuous variable, but our expression (\ref{freeenergy}) is more general.

 \noindent 2. We also note that their expression for the eigenfrequencies (in our notation)
 \begin{equation}
 \omega_\pm^2= k_\perp^2+\left( \sqrt{k_z^2+\frac{\beta^2}{4}} \pm \frac{\beta}{2}\right)^2,
 \end{equation}
 is equivalent to our expression  (\ref{alpha}) in this case,
 \begin{equation}
 \omega^2= {\bf k}^2 \pm \beta \omega \,. \label{77}
 \end{equation}
 This can be seen by solving the quadratic equation (\ref{77}) with $\omega$ as the unknown. The  expression (\ref{77})  was obtained in Refs.~\cite{brevik21a,favitta23} also.
 \noindent 3.  The method of Fukushima {\it et al.} is similar to that of Jiang and Wilczek \cite{jiang19}, and applies primarily  to the case of chiral materials. This is so because the values of $\beta L = \Delta \theta$ for which the important physical effects turn up, are relatively large. Assume for definiteness that $\theta =0$ for $z<0$ so that $\Delta \theta = \theta$  at $z=L$. Then, the appearance of a repulsive force occurs according to these authors at $\theta > 2.38$. This is very much higher than the numbers $ \theta \sim 10^{-18}$ or $ \theta \sim 10^{-2}$ that we have to do with in the High Energy Physics axion case.

 \noindent 4. Our method has allowed us to calculate the temperature-dependent Casimir force between two conducting plates when the axion background has a time derivative $\dot{\theta}=\alpha$ that is uniform and constant. It extends the method and the results we obtained in Ref.~\cite{favitta23}, precisely allowing to calculate the same Casimir force in the high temperature limit. We have shown how in such a case we can have repulsion for the case of high values of the rotation angle $\phi=\frac{1}{2} \sqrt{\frac{\mu}{\varepsilon}} \alpha L$, as happens analogously for chiral and optically active media.
 
  Furthermore, we have discussed the case with $\nabla \theta=\beta$ that is uniform, constant and directed in the normal direction to the plates and we have shown that the zero-temperature Casimir force is analogous to the same one for $\dot{\theta}=\alpha$ that is uniform and constant, while the axionic contribution is suppressed in the high temperature limit.

%\noindent It is important to put together and synthesize the order of magnitude of the axionic correction to the usual Casimir force between the two plates and the validity of our approximation of taking $\alpha$ and $\beta$ to be constant and uniform.
%We have discussed how the case $\beta=0$ cannot be interesting

%However, we leave these cases of interest for future investigations.

\section*{Acknowledgments}
We are grateful to Roberto Passante and Lucia Rizzuto for illuminating discussions and several suggestions.

% Set the ending of a LaTeX document
\end{document}